\documentclass[twocolumn,10pt]{IEEEtran}
\usepackage{flushend}
\usepackage{amsmath}
\usepackage{flushend}
\usepackage{multirow}
\usepackage{color}
\usepackage{enumerate}
\usepackage{easylist}
\usepackage{tikz}
\usetikzlibrary{shapes,arrows,shadows}
\usepackage{times}
\usepackage[symbol]{footmisc}
\usepackage{verbatim}
\usepackage{amsthm}
\usepackage{amssymb }
\usepackage{amsmath}
\usepackage{bm}
\usepackage{cite}
\usepackage{float}
\usepackage{mathrsfs}
\usepackage{amsfonts}
\usepackage{graphicx}
\usepackage{caption}
\usepackage{algorithm, algpseudocode}
\usepackage{subcaption}
\usepackage{footnote}
\usepackage{footmisc}
\usepackage[colorinlistoftodos]{todonotes}
\usepackage{hyperref}
\usepackage{color,soul}
\usepackage{ltablex}
\usepackage{stfloats}
\usepackage{graphicx}

\def\EM{\text{TraHGR}} 
\def\PB{\text{TraHGR-Base}}
\def\PL{\text{TraHGR-large}}
\def\PH{\text{TraHGR-Huge}}
\def\vitF{\text{TNet}} 
\def\vitS{\text{FNet}} 
\def\DB{\text{DB2 ($49$ gestures)}}
\def\EF{\text{DB2-B ($17$ gestures)}}
\def\ES{\text{DB2-C ($23$ gestures)}}
\def\ET{\text{DB2-D ($9$ gestures)}}
\def\C{C}
\def\Po{P_{1}}
\def\Pt{P_{2}}
\def\nina{\text{Ninapro}}

\def\S{S}
\def\W{\mathbf W}
\def\V{\mathbf V}
\def\Q{\mathbf Q}
\def\K{\mathbf K}
\def\Z{\mathbf Z}
\def\P{\mathbf P}
\def\E{\mathbf E}
\def\X{\mathbf X}
\def\N{N}
\def\p{p}
\def\R{\mathbb{R}}
\def\Nd{^{\N\times\d}}
\def\d{D}
\def\w{W}
\def\hsw{^{\S\times\w\times\C}}

\def\nppc{^{\N\times(\Po.\Pt.\C)}}
\def\n1wc{^{\N\times(1.\W.\C)}}
\def\ppcd{^{(\Po.\Pt.\C)\times\d}}
\def\Xp{\X_{\p}}

\def\dh{\d_h}
\def\l{l}
\def\h{h}

\def\L{L}

\def\x{{\mathbf x}}

\def\xp{\x_{\p}}
\def\y{\mathrm{y}}

\algnewcommand\Input{\item[\hspace{6pt}\textbf{Input:}]}
\algnewcommand\Output{\item[\hspace{6pt}\textbf{Output:}]}
\algnewcommand\OutputVal{\textbf{output} }
\newcommand{\etal}{\textit{et al.}}
\def\x{{\mathbf x}}

\title{$\EM$: Transformer for Hand Gesture Recognition via ElectroMyography}

\author{Soheil Zabihi, Elahe Rahimian,  Amir Asif, and Arash Mohammadi
\vspace{-.2in}
\thanks{
S. Zabihi is with Electrical and Computer Engineering, Concordia University, Montreal, QC, Canada, H3G-2W1 (email: s\_zab@encs.concordia.ca). E. Rahimian and A. Mohammadi are with Concordia Institute for Information System Engineering (CIISE), Concordia University, Montreal, QC, Canada, H3G-2W1 (emails: e\_ahimia@encs.concordia.ca, arash.mohammadi@concordia.ca). A. Asif is with Electrical Engineering \& Computer Science, York University, Toronto, Canada, M3J-1P3 (email: asif@eecs.yorku.ca). 
}}

\begin{document}
\date{\today}
\maketitle
\thispagestyle{empty}
\begin{abstract}
Deep learning-based Hand Gesture Recognition (HGR) via surface Electromyogram (sEMG) signals has recently shown considerable potentials for development of advanced myoelectric-controlled prosthesis.   Although deep learning techniques can improve HGR accuracy compared to their classical counterparts, classifying hand movements based on sparse multichannel sEMG signals is still a challenging task. Furthermore, existing deep learning approaches, typically, include only one model as such can hardly maintain acceptable generalized performance in changing scenarios. In this paper, we aim to address this challenge by capitalizing on the recent advances of hybrid models and transformers. In other words, we propose a hybrid framework based on the transformer architecture, which is a relatively new and revolutionizing deep learning model. The proposed hybrid architecture, referred to as the Transformer for Hand Gesture Recognition ($\EM$), consists of two parallel paths followed by a linear layer that acts as a fusion center to integrate the advantage of each module and provide robustness over different scenarios. We evaluated the proposed architecture $\EM$ based on the commonly used second $\nina$ dataset, referred to as the DB2. The sEMG signals in the DB2 dataset are measured in the real-life conditions from $40$ healthy users, each performing $49$ gestures. We have conducted extensive set of experiments to test and validate the proposed $\EM$ architecture, and compare its achievable accuracy with more than five recently proposed HGR classification algorithms over the same dataset.
W have also compared the results of the proposed $\EM$ architecture with each individual path and demonstrated the distinguishing power of the proposed hybrid architecture. The recognition accuracies of the proposed $\EM$ architecture are $86.18\%$, $88.91\%$, $81.44\%$, and $93.84\%$, which are $2.48\%$, $5.12\%$, $8.82\%$, and $4.30\%$ higher than the state-of-the-art performance for $\DB$, $\EF$, $\ES$, and $\ET$, respectively.
\end{abstract}
\begin{IEEEkeywords}
Electromyogram (EMG), Deep Neural Networks (DNNs), Machine Learning (ML), Transformers, Prosthetic, Classification, Hand Gesture.
\end{IEEEkeywords}
\vspace{-.1in}
\section{Introduction}\label{sec:Introduction}
To improve the quality of life of people with upper limb amputation, recently, there has been a growing interest in development of learning-based myoelectric prostheses using multi-channel surface Electromyogram (sEMG) signals~\cite{2_Dario, Dario,3_Dario}. The information obtained from the sEMG signals, which are related to the neural activities of the underlying muscles, is used to decode the movements of the targeted limb. Generally speaking, sEMG signals can be collected via two different recording techniques, i.e., sparse multi-channel sEMG, and High-Density sEMG (HD-sEMG). The former, typically, consists of a limited number of electrodes with sparse placement, which is the commonly used sEMG recording modality in wearable systems~\cite{Ergeneci, WeiNet}. On the other hand, a HD-sEMG device consists of a grid of electrodes that collect information about the temporal and spatial electrical activities of the underlying muscles enabling them to capture large amount of data~\cite{WeiNet,YuNet,GengNet}. This, however, results in increased complexity of the underlying system, which in turn challenges their ease of applicability in wearable systems and adds latency to
the processing pipeline. Recent studies~\cite{TNSRE_Elahe},therefore, focus on the use of sparse multi-channel sEMG devices given their ease of use and reduced complexity~\cite{Farina14}. The use of sparse multi-channel sEMG datasets can, however, challenge the gesture recognition performance due to its sensitivity  to the electrode location. In particular, while Deep Neural Networks (DNNs) achieve high performance in gesture recognition using HD-EMG devices, their efficacy is limited for sparse multi-channel sEMG devices in which a shallow dataset is collected with a lower sampling rate and limited number of electrodes. Inspired by the above discussion, as well as the significant gap between the performance of existing methods for HD-EMG and the sparse multi-channel sEMG approaches, the paper aims to present a new hybrid learning framework for achieving superior performance using sparse multi-channel signals.

Generally speaking, the developed methods for classifying hand movements can be classified into the following three main categories: (i) Traditional approaches based on Machine Learning (ML) architectures~\cite{LDA-SVM, SVM, LDA, DB5, AtzoriNet}; (ii) DNN-based techniques~\cite{FD-TBME, JMRR_Elahe, Globalsip_Elahe, Icassp_Elahe,ICASSP2_Elahe, pattern_letter2019, WeiNet, AtzoriNet}, and; (iii) Hybrid methodologies~\cite{YuNet} that combine multiple models. The common approach to perform Hand Gesture Recognition (HGR) in traditional methods (Case (i)) is to extract hand-crafted (engineered) features to train classical ML models such as Linear Discriminant Analysis (LDA), Support Vector Machine (SVM), and Random Forests (RF). In recent years, there has been a growth of interest in developing deep learning-based approaches (Case (ii)) for HGR, which show encouraging classification results~\cite{Wei2017, YuNet}. In particular, deep learning techniques provide an effective venue to automatically extract features from sEMG data and improve gesture recognition accuracy compared to their classical counterparts. However, many of the existing deep learning approaches involve only a single model, which can hardly maintain acceptable generalization performance in different scenarios, especially when configured for a specific scenario. The paper addresses this gap and focuses on the design of a transformer-based hybrid solution (Case (iii)) that has  potentials for extracting temporal and spatial properties and improving HGR accuracy. In particular, transformer neural networks~\cite{Vaswani}, which are a relatively new and revolutionizing deep learning model, were first used for Natural Language Processing (NLP) tasks~\cite{Bert}. Recently, they have been used to improve the performance of several other applications such as Computer Vision (CV)~\cite{ViT} and speech recognition tasks~\cite{EEG1}. Inspired by the recent advances of transformers in the various domains, in this paper, we aim to leverage the power of transformers to increase the HGR task performance.

\vspace{.1in}
\noindent
\textbf{Contributions}: Motivated by the recent accomplishments of hybrid deep learning approaches, the paper proposes a hybrid DNN framework based on the transformer for the HGR task. Referred to as the Transformer for Hand Gesture Recognition ($\EM$),  the proposed framework consists of two parallel paths (one Temporal transformer Network ($\vitF$) and one Feature transformer Net ($\vitS$)) followed by a linear layer, which integrates the output of each path to provide robustness across different scenarios. The $\vitF$ is used to extract temporal features while simultaneously the $\vitS$ is utilized to extract spatio-temporal features, which are then fused to augment the discriminating power of the model and improve the overall performance of HGR classification task.
Performance of the proposed $\EM$ framework is evaluated using the second $\nina$ database~\cite{1_Ninapro, 2_Ninapro, 3_Ninapro} referred to DB2, which is a commonly used dataset that provides sparse multi-channel sEMG signals from various hand movements similar to those obtained in real-life conditions. Thus, $\nina$ dataset enables  development of innovative DNN-based recognition solutions for HGR task. We have conducted extensive set of experiments to test and validate the proposed $\EM$ architecture, and compare its achievable accuracy with more than five recently proposed HGR classification algorithms based on the same datasets. Results show that the proposed $\EM$ framework provides superior performance over all its counterparts. More specifically, the DB2 dataset has divided into $3$ exercises with different hand gestures; i.e., $\EF$, $\ES$, and $\ET$. Our proposed architecture $\EM$ outperforms all existing solutions over the DB2 dataset and its sub-exercises.
In summary, the contributions of the paper are as follows:
\begin{itemize}
\item  The paper for the first time, to the best of our knowledge, develops a hybrid Transformer based architecture for the task of HGR via sparse multi-channel sEMG.
\item The paper demonstrates the superior performance of the propose hybrid architecture $\EM$ and its ability to extract more distinctive information for gesture recognition over the single models; i.e., $\vitF$ and $\vitS$.
\item The paper analyzes the effect of different  window sizes $200$ms, $150$ms, and $100$ms on overall performance and the model complexity.
\item The paper classifies a high number (49) of gestures with a high accuracy. More specifically, compared with the proposed architectures in the recent state-of-the-art studies, $\EM$ improves the recognition accuracies to $86.18\%$ on $\DB$, to $88.91\%$ on the $\EF$, to $81.44\%$ on  $\ES$, and to $93.84\%$ on the $\ET$.
\end{itemize}
The rest of the paper is organized as follows: In Section~\ref{sec:RelWorks}, an overview of related works is provided. Section~\ref{sec:database} describes the database and pre-processing step. In Section~\ref{sec:model}, we present details of the proposed $\EM$ architecture. The experiments and results are presented in Section~\ref{sec:results}. Finally, the conclusion is presented in Section~\ref{sec:con}.

\vspace{-.1in}
\section{Related Works}\label{sec:RelWorks}
The existing researches on prosthetic myoelectric control focus primarily on traditional ML approaches as a common strategy for HGR~\cite{Cote}. In such methods, handcrafted features, in time domain, frequency domain, or time-frequency domain~\cite{YuNet}, are first extracted by human experts, which are then fed to a ML classifier. Extraction and feature selection, however, can affect the overall performance~\cite{Rami}, as such some researchers~\cite{WeiNet} have explored and integrated several classical feature sets that provide multi-view of sEMG signals to achieve higher gesture recognition accuracy. On the other hand, different classifiers such as SVM, LDA, RF, and Principal Components Analysis (PCA) are utilized in the literature~\cite{AtzoriNet, ZhaiNet, DB5, 1_Ninapro, Rami} to increase the discriminating power of the model and improve gesture recognition performance.

Although the traditional ML-based approaches have shown strong potential for HGR task, more recently, there has been a great deal of interest in using deep-learning architectures to process multi-channel sEMG signals and increase the discrimination power of the model. In particular, it has been shown~\cite{AtzoriNet} that the automatic feature extraction used in deep learning architecture can lead to higher classification accuracy. More specifically, the authors in~\cite{AtzoriNet}, for the first time, used the Convolutional Neural Network (CNN) architecture to classify hand gestures, which showed its potential to improve the overall performance compared to existing traditional approaches. This achievement was the starting point for considering CNN as a promising approach in the context of sEMG data classification~\cite{pattern_letter2019, DingNet, Icassp_Elahe}. For example, in Reference~\cite{GengNet}, the authors proposed and used the CNN architecture to extract spatial information from sEMG signals and perform HGR classification. In addition to CNN-based architectures, some researches~\cite{Quivira, Atashzar} used Recurrent Neural Networks (RNNs) to extract the temporal features from the sEMG signals. RNNs are used because sEMG signals are sequential in the nature, and recurrent-based networks such as Long Short Term Memory (LSTM) can extract the patterns in a sequence of sEMG data treating HGR as a sequence modeling task. In addition, it has been shown~\cite{TCN} that the proper design of CNN architectures can outperform RNNs in sequence modeling. In this respect, some researchers~\cite{Globalsip_Elahe, TNSRE_Elahe, ICASSP2_Elahe} used temporal convolutions for HGR and showed its potentials to extract temporal features.

Alternatively, hybrid architectures such as CNN-RNN have shown promising results in classifying hand movements~\cite{JMRR_Elahe, YuNet}, as they benefit from advantages of different modules in extracting temporal and spatial features. Meanwhile, with the advent of the attention mechanism~\cite{Vaswani}, transformers are being considered as a new ML technique for sequential data modeling~\cite{ECG,EEG2}. Capitalizing on the recent success of transformers in various fields such as machine translation~\cite{GPT3, Bert}, speech recognition~\cite{acoustic} and computer vision~\cite{ViT}, we aim to examine its applicability and potentials for sEMG-based gesture recognition. In other words, we recognized an urgent need to develop a transformer-based hybrid architecture to augment the recognition accuracy of HGR. In this paper, we introduce the $\EM$ architecture, which increases the accuracy of sEMG decoding in the classification of human hand movements. In addition, we examine the complexity and performance of different types of $\EM$ architectures. 

\vspace{-.1in}
\section{Material and Methods}\label{sec:database}
\subsection{Database}\label{sec:dataset}
The proposed $\EM$ architecture is evaluated on the second $\nina$ database~\cite{1_Ninapro, 2_Ninapro, 3_Ninapro}, which is a publicly available dataset for sEMG-based human-machine interfacing. The second database $\nina$, which is referred to as the DB2, was collected from $40$ users. Each user performs $49$ movements in which each movement is repeated $6$ times, each time lasting for $5$ seconds, followed by $3$ seconds of rest. The sEMG signals were gathered using Delsys Trigno Wireless EMG system with $12$ wireless electrodes, sampled at $2$ kHz. The DB2 dataset was presented in three exercises B, C, and D, which consist of different types of movements. In particular, Exercise B, C, and D consists of $17$, $23$, and $9$ movements, respectively. For the rest of this paper,  Exercise B, C, and D are referenced to $\EF$, $\ES$, and $\ET$, respectively. We followed the recommendations provided by previous studies~\cite{1_Ninapro, 2_Ninapro, 3_Ninapro} and consider the repetitions $1$, $3$, $4$, and $6$ of each movement as the training set, and the remaining repetitions, i.e., $2$ and $5$,  as the test set.

\subsection{Pre-processing Step}\label{sec:pre-process}
The EMG signals are pre-processed for classification purposes before being fed into the proposed architecture. The pre-processing phase consists of three steps, i.e., low-pass filtering, normalization, and segmentation. More specifically, we followed the procedure outlined in previous studies~\cite{GengNet, Wei2017, AtzoriNet} and used the low-pass Butterworth filter. To enhance the performance of the proposed architecture, we applied the low-pass filter three times with different order of filters, namely $1$, $3$, and $5$, and then concatenated all filtered signals together to form three-channel sEMG signals. Then, for the normalization step, we are inspired by the  $\mu$-law normalization technique introduced in~\cite{Icassp_Elahe, TNSRE_Elahe}, which is defined as follows
\begin{equation}\label{mu_law}
F(x_t) = \text{sign}(x_t)\frac{\ln{\big(1+ \mu |x_t|\big)}}{\ln{\big(1+ \mu \big)}},
\vspace{-.1in}
\end{equation}
where the input scaler is represented by $x_t$, and the new range of the signal is indicated by parameter $\mu$. After normalization, the sEMG signals are segmented with a sliding window. More specifically, a common strategy used in previous studies is to segment sEMG signals with a window length that should be less than $300$ms to meet an acceptable delay time~\cite{24} for practical applications. Therefore, in this study, the results are reported with three sliding windows of different lengths, i.e., $200$ms, $150$ms, and $100$ms, each with a step size of $10$ms. Each input from the sEMG signal segmentation phase is denoted by $\X  \in \R\hsw$, where $\S$ shows the number of sensors in the DB2 dataset, $\w$ shows the number of samples of electrical activities of muscles obtained at the rate of $2$ kHz for a window of $200$ms, $150$ms, or $100$ms, and $\C$ denotes the number of channels of the sEMG signals. This completes description of the pre-processing procedure. Next, the proposed $\EM$ architecture is presented.

\section{The $\EM$ Framework}\label{sec:model}
In this section, we explain details of the proposed $\EM$ architecture for HGR. The $\EM$ architectural design is based on transformers in which the attention mechanism is employed. The attention mechanism has been used in previous studies~\cite{YuNet, TNSRE_Elahe, ICASSP2_Elahe} in conjunction with CNNs and/or recurrent-based architectures for HGR task. However, in this paper, we show that transformer-based architectures that rely solely on attention mechanisms can perform better than previous studies in which CNN, RNN, and hybrid architectures (e.g., attention-based hybrid CNN-RNN) have been adopted. The overall proposed architecture is illustrated in Fig.~\ref{arc}, which is inspired by the Vision Transformer (ViT)~\cite{ViT}, in which each input is divided into patches, and the network is supposed to perform label prediction based on the sequence of these patches. As shown in Fig.~\ref{arc}, the proposed $\EM$ consists of a $\vitF$ path implemented in parallel with a $\vitS$ path followed by a linear layer, which acts as the fusion centre combining the extracted features from each of the two parallel paths in order to classify the hand gestures. In the following, we will further elaborate on the details of the proposed architecture.

\subsection{Embedded Patches}\label{sec:embed}
In this sub-section, we focus on the input of the transformer encoder, which is a sequence of embedded patches. As illustrated in Fig.~\ref{arc}, the embedded patches are constructed from patch embeddings and position embeddings, which are described below.

\subsubsection{Patch Embeddings}\label{sec:patch}
As mentioned earlier, we split each segment of sEMG signals $\X$ into non-overlapping patches $\Xp=\{\x^{i}_p\}^{N}_{i=1}$.  More specifically, each segment $\X\in\R\hsw$ is divided into $\N$ non-overlapping patches in which each patch is flattened. We represented the sequence of these flatten patches with $\Xp \in \R\nppc$, where $\C$ denotes the number of channels, ($\Po$, $\Pt$) shows the size of each patch, and $\N = \S.\w / \Po.\Pt$ represents the length of this sequence, i.e., the number of patches. As shown in Fig.~\ref{arc}, we applied two types of patching:
\begin{itemize}
\item \textit{Temporal Patching:} Here, the size of each patch is ($1$, $\w$); therefore, the number of patches is $\N = \S$. This type of patching is called Temporal Patching because each patch contains the information from only one of the sensors in the dataset for a sequence with length $\w$.  The $\vitF$ path is designed in such a way that they can take into account the temporal patches as the input. 
\item \textit{Featural Patching:} We set the size of each patch to ($\S$, $\S$), i.e., $\Po=\Pt=\S$, therefore, the number of patches is $\N = \w / \S$. We refer to this type of patching as Featural because each patch contains the information of all $\S$ sensors for a sequence with length of $\S$. Therefore, both spatial and temporal information are included in a Featural patch. The Featural patches are provided as the input only to the $\vitS$ layer as shown in Fig.~\ref{arc}.   
\end{itemize}
Finally, a linear mapping is introduced to create the embedding for each of these patches (Fig.~\ref{arc}). More specifically, a matrix $\E\in\R\ppcd$ is shared among different patches to linearly project  each patch into the model dimension $\d$ (Eq.~\eqref{eq:patch}). The output of this projection is known as the Patch Embeddings.

\subsubsection{Class Token}\label{sec:class}
Similar to the Bert framework~\cite{Bert}, a trainable embedding is prepended to the sequence of patch embeddings ($\Z_0^0 = \x_{\text{class}}$) with the goal of capturing the meaning of the entire segmented input as a whole. More specifically, the class token's embedding after the last transformer encoder layer ($\Z_L^0$) is used for classification purposes (Eq.~\eqref{eq:out}).

\subsubsection{Position Embeddings}\label{sec:position}
\begin{figure*}[t!]
\centering
\includegraphics[scale=.26]{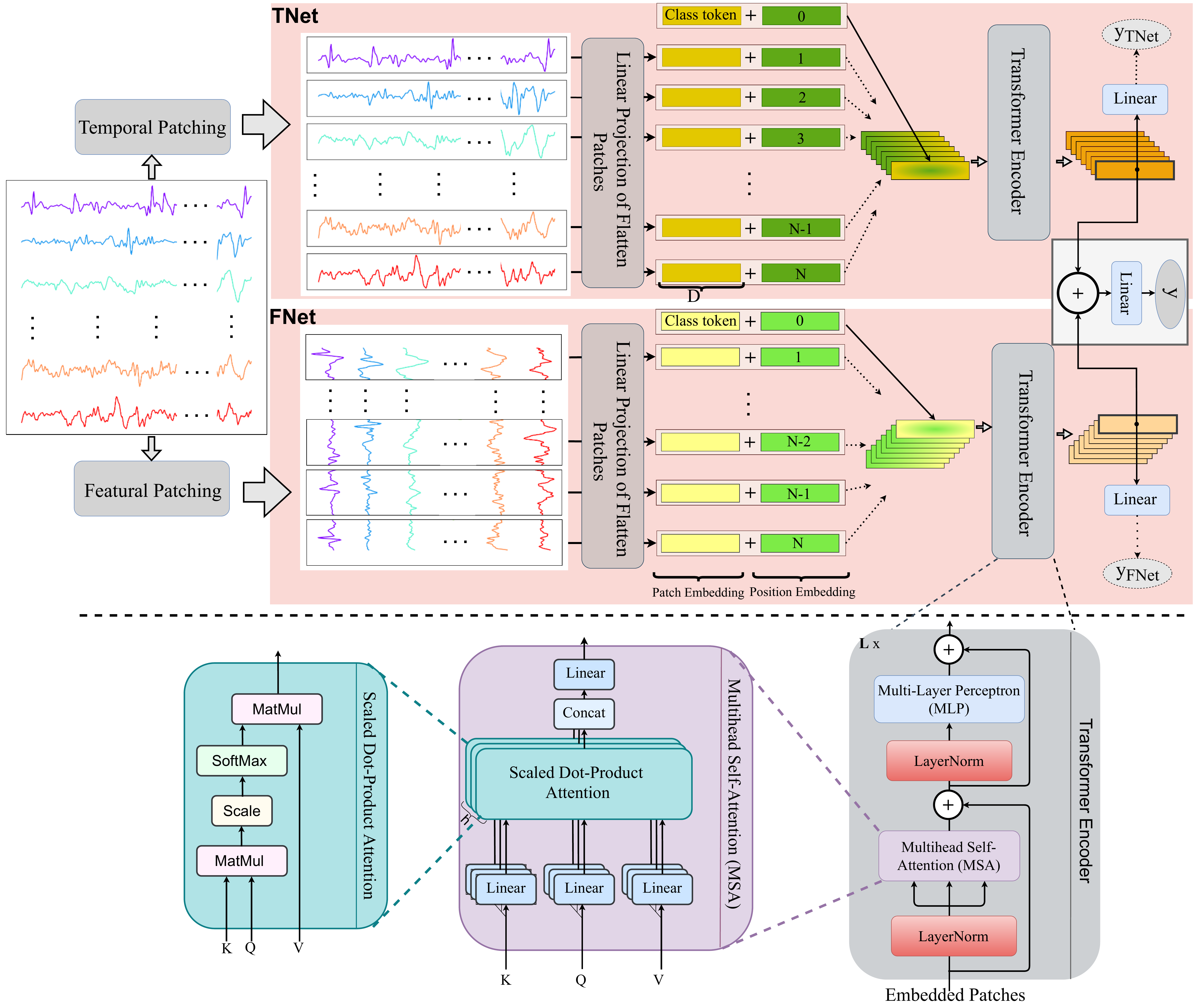}
\vspace{-.1in}
\caption{\small \textbf{The proposed $\EM$ architecture} consists of two parallel paths (one $\vitF$ and one $\vitS$). Each segment of sEMG signals $\X$ is divided into $\N$ non-overlapping patches.  The $\EM$ uses the $\vitF$ path to get the temporal patches while simultaneously the $\vitS$ is utilized to consider the featural patches. In both $\vitF$ and $\vitS$, the patches are mapped linearly into the model dimension $\d$. We refer to the output of this step as ``Patch Embedding''. Then, a ``class token'' is prepended to the sequence of patch embeddings which is finally used for the classification purpose. The ``Positional Embedding'' is added to the ``Patch Embedding'' to retain the positional information. The output of this step is called ``Embedded Patches'' and is fed to the Transformer encoder consisting of $\L$ layers,  each layer consisting of MSA and MLP modules. Finally, we add the output of the $\vitF$ and $\vitS$ class tokens to get the final representation, which then acts as the input to the linear layer.  \label{arc}}
\end{figure*}

As HGR based on sEMG signals is a time-series processing task, the order of data is an essential part for  sequence modeling. Recurrent-based architectures such as LSTM inherently consider signal order,  however, transformers do not process the input sequentially and combine the information of all the elements through attention mechanism. Therefore, there is a need to encode the order of each element in the sequence. This is where positional embedding comes in. In fact, position embedding allows the network to  determine  where a particular patch came from. There are several ways to retain position information at the transformer input, e.g., Sinusoidal positional embedding, 1-dimensional positional embedding, 2-dimensional positional, and Relative positional embeddings embedding~\cite{acoustic, ViT}. Following~\cite{ViT}, we used the standard trainable 1-dimensional positional embeddings. As shown in Fig.~\ref{arc}, position embeddings indicated by $\E_{pos}\in\R^{(\N + 1)\times\d}$ is added to the patch embeddings. The formulation which governs patch and position embeddings is as follows
\begin{eqnarray}
\Z_0 = [\x_{\text{class}}; \xp^1\E; \xp^2\E;\dots; \xp^\N\E] + \E_{pos}. \label{eq:patch}
\end{eqnarray}
The output of Eq.~\eqref{eq:patch} is called Embedded Patches, which are fed as an the input to the transformer encoder.

\subsection{Transformer Encoder}\label{sec:transformer}
The transformer encoder takes the $\Z_0$ as an input. This block is inspired from the main transformer encoder introduced in~\cite{Vaswani}, which treats all embedded patches as tokens. As illustrated in Fig.~\ref{arc}, the transformer encoder consists of $\L$ layers. Each layer contains two modules, namely the Multihead Self-Attention (MSA) and a Multi-Layer Perceptron (MLP) module, i.e.,
\begin{eqnarray}
\begin{aligned}
\Z^{'}_l &=& \textit{MSA}(\textit{LayerNorm}(\Z_{\l-1})) + \Z_{\l-1}, &&& \l = 1\dots\L\label{eq:MSA}
\end{aligned}
\end{eqnarray}
\vspace{-.33in}
\begin{eqnarray}
\begin{aligned}
\Z_l &= \textit{MLP}(\textit{LayerNorm}(\Z^{'}_{\l})) + \Z^{'}_{\l}, &&&&&&& \l = 1\dots\L \label{eq:MLP}
\end{aligned} 
\end{eqnarray}
 It is worth noting that a layer-normalization~\cite{layernorm} is used before MSA and MLP modules, and the residual connections are applied to address degradation problem. 
The MLP module consists of two linear layers in which the first layer is followed by Gaussian Error Linear Unit (GELU) activation function. Moreover, the MSA module is defined based on the Self-Attention (SA) mechanism, which is discussed next.

\subsubsection{Self-Attention (SA)}\label{sec:sa}
The SA mechanism~\cite{Vaswani} measures the pairwise similarity of each query and all the keys and obtains a weight for each value. Finally, the output is computed based on the weighted sum over of all values. In particular, if we define an input $\Z \in \R\Nd$ consisting of $\N$ vectors, each of length  $\d$, the three matrices, i.e., Queries $\Q$, Keys $\K$, and Values $\V$, are calculated as follows
\begin{eqnarray}
[\Q, \K, \V]  = \Z\W_{QKV}\label{eq.2},
\end{eqnarray}
where $\W_{QKV} \in \R^{\d\times 3\dh}$ denotes the trainable weight matrix and $\dh$ shows the length of each vector in $\Q$, $\K$, and $\V$. To measure the weights for $\V$, the dot-product of $\Q$ and $\K$ is calculated, then scaled with $\sqrt{\dh}$. These weights are converted to the probabilities $\P \in \R^{\N\times \N}$ using the softmax function as follows
\begin{eqnarray}
\begin{aligned}
\P = \textit{softmax}(\frac{\Q\K^T}{\sqrt{\dh}}).\label{eq.3}
\end{aligned}
\end{eqnarray}
Finally, he output of SA mechanism is computed as follows
\begin{eqnarray}
SA(\Z) = \P\V \label{eq.4}.
\end{eqnarray}
By using the attention mechanism, the model pinpoints a specific information in the input sequence.

\subsubsection{Multihead Self-Attention (MSA)}\label{sec:msa}
Here, the SA mechanism is used for $\h$ times in parallel, allowing the architecture to pinpoint specific pieces of information in the input sequence for each head differently. In particular, each head has its own trainable weight matrix. The final matrix in the MSA mechanism is a projection of the concatenated outputs of the $\h$ heads, which is formulated as follows
\begin{eqnarray}
MSA(\Z) = [SA_1(\Z); SA_2(\Z); \dots; SA_h(\Z)]\W_{MSA},\label{eq:MSA}
\end{eqnarray}
where $\W_{MSA} \in \R^{\h.\dh \times \d}$. Here, $\dh$ is set to $\d / \h$ to keep the number of parameters constant when $\h$ changes.

\subsection{$\EM$'s Output}\label{sec:output}
As shown in Fig.~\ref{arc}, the $\EM$ consists of two paths, i.e., $\vitF$ and $\vitS$. For each path, the aforementioned calculations (Eqs.~\eqref{eq:patch}-\eqref{eq:MSA}) are performed in parallel. Then, the predicted class labels of each path is calculated based on its corresponding $\Z_L^0$ as follows
\begin{eqnarray}
\y_{\textit{path}} = \textit{Linear}(\textit{LayerNorm}(\Z_{\L}^0)_{\textit{path}}), \label{eq:out1}
\end{eqnarray}
where $\textit{path}\in\{\vitF, \vitS\}$. Finally, the output of the $\EM$ is calculated based on the sum of $\Z_L^0$ in the $\vitF$ and $\vitS$ as follows
\begin{eqnarray}
\y = \textit{Linear}(\textit{LayerNorm}[(\Z_{\L}^0)_{\vitF} + (\Z_{\L}^0)_{\vitS}]).\label{eq:out}
\end{eqnarray}
It is worth mentioning that $\y_{\vitF}$, $\y_{\vitS}$, and $\y$ are used for $\EM$ training. More details are provided in the subsection~\ref{sec:loss1}. This completes description of the proposed $\EM$ architecture, next, its performance is evaluated through several experiments.

\begin{table*}[t!]
\centering
\renewcommand\arraystretch{1.7}
\caption{\small Descriptions of $\EM$ architecture variants, $\vitF$, and $\vitS$ for $\DB$. Number of parameters (Params) are reported for window sizes $200$ms, $150$ms, and  $100$ms.}
\label{table1}
{\begin{tabular}{ c c c c c c c c}
\hline
\hline
\multicolumn{1}{c}{\multirow{2}[2]{*}{\textbf{Model}}}
& \multicolumn{1}{c}{\multirow{2}[2]{*}{\textbf{Layers ($\L$)}}}
& \multicolumn{1}{c}{\multirow{2}[2]{*}{\textbf{Model dimension ($\d$)}}}
& \multicolumn{1}{c}{\multirow{2}[2]{*}{\textbf{MLP size}}} 
& \multicolumn{1}{c}{\multirow{2}[2]{*}{\textbf{Number of heads ($\h$)}}} 
&  \multicolumn{3}{c}{\multirow{1}[2]{*}{\textbf{Params}}}
\\
\cline{6-8}
&
&
&
&
& \textbf{$200$ms}
& \textbf{$150$ms}
& \textbf{$100$ms}
\\
\hline
\textbf{$\PB$}
& 1
& 32
& 128
& 4
& 83,731
& 74,259
& 63,603
\\
\textbf{$\PL$}
& 2
& 64
& 256
& 4
& 316,051
& 297,107
& 275,795
\\
\textbf{$\PH$}
& 1
& 144
& 720
& 8
& 846,579
& 803,955
& 756,003
\\
\textbf{$\vitF$}
& 1
& 144
& 720
& 8
& 472,513
& 431,041
& 384,385
\\
\textbf{$\vitS$}
& 1
& 144
& 720
& 8
& 366,673
& 365,521
& 364,225
\\
\hline
\hline
\end{tabular}}
\end{table*}
\begin{table}[t!]
\centering
\renewcommand\arraystretch{1.7}
\caption{\small The average of gesture recognition accuracy over all subjects using different window size ($200$ms, $150$ms, and $100$ms) for $\EM$ architectures variants, $\vitF$, and $\vitS$ for $\DB$. \label{table2}}
{\begin{tabular}{c c c c}
\hline
\hline

\multicolumn{1}{c}{\multirow{2}[2]{*}{\textbf{Model}}}
&\multicolumn{3}{c}{\multirow{1}[2]{*}{{\footnotesize \textbf{Accuracy $\pm$ STD}}}}
\\
\cline{2-4}

& \textbf{$200$ms}
& \textbf{$150$ms}
& \textbf{$100$ms}
\\
\hline
\multicolumn{1}{c}{\textbf{$\PB$}} & $78.60$ $\pm$ $6.03$  &  $77.54$ $\pm$ $5.99$ &  $76.17$ $\pm$ $6.09$
\\

\multicolumn{1}{c}{\textbf{$\PL$}} & $83.58$ $\pm$ $5.48$ & $82.58$ $\pm$ $5.60$ & $81.30$ $\pm$ $5.87$
\\

\multicolumn{1}{c}{\textbf{$\PH$}} & $\textbf{86.18}$ $\pm$ $\textbf{4.99}$ & $\textbf{85.43}$ $\pm$ $\textbf{5.24}$  & $\textbf{84.13}$ $\pm$ $\textbf{5.21}$
\\

\multicolumn{1}{c}{\textbf{$\vitF$}} & $83.39$ $\pm$ $5.44$ &  $82.81$ $\pm$ $5.60$ &  $81.43$ $\pm$ $5.88$
\\

\multicolumn{1}{c}{\textbf{$\vitS$}} & $80.72$ $\pm$ $5.82$ & $80.05$ $\pm$ $6.03$  & $79.38$ $\pm$ $6.15$
\\
\hline
\hline
\end{tabular}}
\end{table}

\section{Experiments and Results}\label{sec:results}
In this section, we evaluate performance of the proposed $\EM$ architecture through a series of experiments. In all experiments, the Adam optimizer~\cite{adam} was used with learning rate of $0.0001$ and the weight decay of $0.001$. Moreover, the batch size is set to $512$. Table~\ref{table1} shows different configurations of the hyperparameters in the $\EM$ architecture resulting in different variants of the model denoted by $\PB$, $\PL$, and $\PH$. These variants are then used for training and evaluation purposes with window size $\in \{$200$ms, $150$ms, $100$ms\}$. Moreover, we evaluated the performance of a single deep model ($\vitF$ or $\vitS$) when they are trained independently.  In Table~\ref{table1}, the number of parameters (Params) is calculated for $\DB$ while this number will be less for $\EF$, $\ES$, and $\ET$.

\subsection{Loss Function}\label{sec:loss1}
The loss function $\mathcal{L}$ of the $\EM$ consists of the following three components 
\begin{eqnarray}
\mathcal{L} = \mathcal{L}_{\vitF} + \mathcal{L}_{\vitS} + \mathcal{L}_{\EM} \label{eq:loss},
\end{eqnarray}
where the first term $\mathcal{L}_{\vitF}$ is loss of the $\vitF$ path in the proposed $\EM$ architecture. More specifically, cross-entropy loss is considered for measuring classification performance using the $\vitF$'s output $\y_{\vitF}$ (Eq.~\eqref{eq:out1}) and the target values. Similarly, the second term $\mathcal{L}_{\vitS}$ is the cross-entropy loss computed using the second path ($\vitS$) of the $\EM$ architecture where $\vitS$'s outputs $\y_{\vitS}$  (Eq.~\eqref{eq:out1}) are considered. Finally, the last term $\mathcal{L}_{\EM}$ is calculated using the $\EM$'s output $\y$ (Eq.~\eqref{eq:out}).

\subsection{Evaluation of the Proposed $\EM$ Architecture}\label{sec:eval}
This subsection provides evaluations on the prediction performance of the proposed hybrid transformer-based architecture. In this regard, first, we compare different variants of the $\EM$ architecture and show the effect of different hyperparameters (e.g., number of layers, model dimension, MLP size, and number of heads) on the overall accuracy. Then, to demonstrate the performance of the hybrid transformer, we also compare the $\EM$ architecture with single deep models, i.e., $\vitF$ and $\vitS$.

Table~\ref{table2} shows recognition accuracy, which is averaged over all subjects for the test set. From Table~\ref{table2}, it can be observed that the proposed $\PH$ architecture outperformed other $\EM$ architecture variants ($\PB$ and $\PL$) when evaluated based on the $\DB$ for the same window size. However, as shown in Table~\ref{table1}, the number of parameters of the $\PH$  is much higher than that of the $\PB$ and $\PL$ models. This fact indicates that increasing the number of layers ($\l$), model dimension ($\d$), MLP size, and number of heads ($\h$) have a positive effect on the model's accuracy, however, this comes with the cost of increasing the complexity. In addition, as shown in Table~\ref{table1}, each model has a larger number of trainable parameters for window size $200$ms than its counterpart in the window size of $150$ms or $100$ms, resulting in a higher complexity. However, as shown in Table~\ref{table2}, larger window size can further improve the results because the transformer in the proposed model has access to a longer sequence length.

\begin{figure*}[t!]
\centering
\includegraphics[scale=.18]{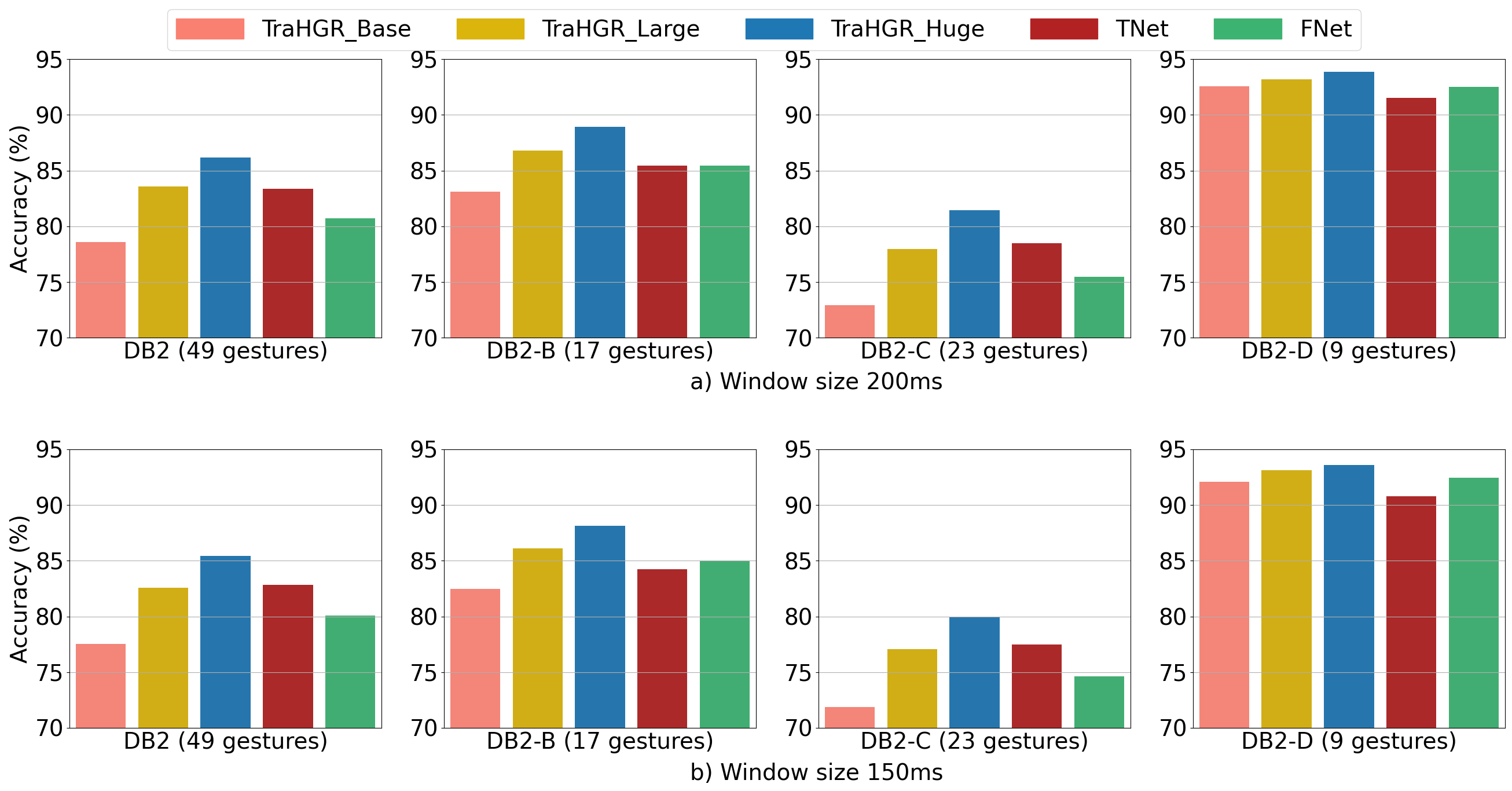}
\vspace{-.05in}
\caption{\small Breakdown of $\DB$ performance in $\EF$, $\ES$, and $\ET$ exercises.}\label{fig:bar200}
\end{figure*}

We also trained and evaluated the proposed model for $\EF$, $\ES$, and $\ET$, independently. In Fig.~\ref{fig:bar200},  performance of the proposed architectures for $\DB$ and its breakdown into three corresponding exercises, B, C, and D are shown. It can be observed that for both window sizes of $200$ms and $150$ms achieving a high accuracy for $\ES$ is a challenging task, while the model has achieved a high accuracy for $\EF$ and $\ET$. Moreover, as shown in Table~\ref{table2} and Fig.~\ref{fig:bar200}, the proposed hybrid architecture $\PH$ is more accurate than the single deep models ($\vitF$ and $\vitS$). Therefore, the hybrid approach integrates advantages of two parallel paths through their integration such that the results can be aggregated to improve the predictive performance. It is worth mentioning that for hybrid models such as $\PB$, $\PL$, and $\PH$, the classification accuracy is calculated using the output of Eq.~\eqref{eq:out}, while for single deep models such as $\vitF$ and $\vitS$ this number is computed using the output of Eq.~\eqref{eq:out1}.

\subsection{Statistical Analysis}\label{sec:stat}
\begin{figure}[t!]
\centering
\includegraphics[scale=.48]{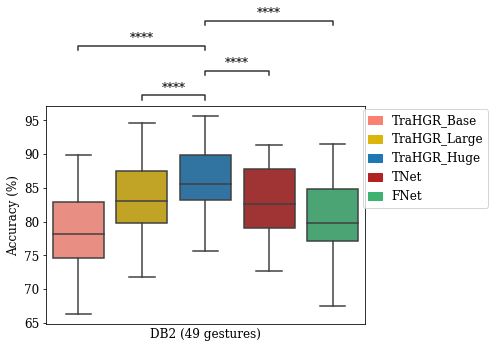}
\vspace{-.1in}
\caption{\small The accuracy boxplots for all $\EM$ architecture variants, $\vitF$, and $\vitS$ for all $49$ gestures in dataset $\nina$ DB2. The IQR of each model is shown by a boxplot for all users. The Wilcoxon signed-rank test is used to compare the $\PH$ with other architectures; i.e., $\PB$, $\PL$, $\vitF$, and $\vitS$ (ns: $0.05 < \text{p} \leq 1$, $^{*}: 0.01 < \text{p} \leq 0.05$, $^{**}: 0.001 < \text{p} \leq 0.01$, $^{***}: 0.0001 < \text{p} \leq 0.001$, $^{****}: \text{p} \leq 0.0001$).}\label{boxplot}
\end{figure}

For the window size $200$ms, we performed statistical analysis on the effectiveness of the observations for $\DB$. Therefore, following~\cite{transfer-learning, TNSRE_Elahe}, we considered each user as a separate dataset and conduct Wilcoxon signed-rank test~\cite{Wilcoxon}. According to the results shown in Fig.~\ref{boxplot}, the difference in accuracy between $\PH$ and other proposed architectures such as $\PB$, $\PL$, $\vitF$, and $\vitS$, for window sizes $200$ms were considered statistically significant by the Wilcoxon signed-rank test. In Fig.~\ref{boxplot}, a $p$-value is annotated by the following markers: (i) $0.05<\text{p-value} \leq1$ is marked as not significant (ns); (ii) $0.01 < \text{p-value} \leq 0.05$ is depicted with $*$; (iii) $0.001< \text{p-value} \leq 0.01$ is marked as $**$; (iv) $0.0001 < \text{p-value} \leq 0.001$ is shown with $***$; and (v) $\text{p-value} \leq 0.0001$ is marked with $****$. In Fig.~\ref{boxplot}, for each proposed model, the performance distribution over all users is illustrated. More specifically, each boxplot shows the Interquartile Range (IQR), which presents the performance of each model for all users into quartiles. The horizontal line in each boxplot illustrates the median performance.

\subsection{Position-Wise Cosine Similarity}\label{sec:pos}
\begin{figure*}[t!]
\centering
\includegraphics[scale=.55]{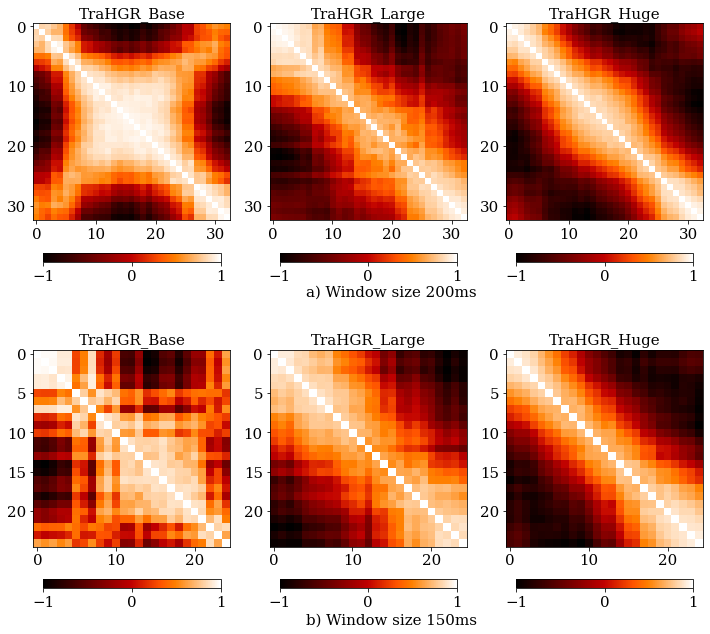}
\vspace{-.1in}
\caption{\small Position embedding similarities for $\vitS$ path in $\PB$, $\PL$, and $\PH$ architectures: (a) window size is $200$ms, and (b) window size is $150$ms. Each row in each figure represents the cosine similarity between one embedding position and all the other embeddings. Brighter in the figures indicates more similarity.}\label{fig:pos}
\end{figure*}

In Fig.~\ref{fig:pos}(a) and (b), the position-wise cosine similarities between one position embedding and other embeddings are depicted for window sizes of $200$ms and $150$ms, respectively. In particular, close positions are brighter, that is, more similar, resulting in a matrix in which the main  diagonal and its neighbors have the brightest colors. As shown in Fig.~\ref{fig:pos}, for both window sizes $200$ms and $150$ms, $\PH$ captures the position meanings better than $\PL$, and $\PL$ better than $\PB$. As a result, we conclude that a more complex architecture can strengthen position embedding and include more location information for transformer encoders. Moreover, as shown in Fig.~\ref{fig:pos}, for longer window sizes, the sequential nature of sEMG signals can be better encoded. For instance, as shown in Fig.~\ref{fig:pos}(b), the position embeddings for $\PB$ architecture did not learn the meaning of positions completely. As a result, it can be concluded that the window size has a direct effect on obtaining position information for the transformer encoder.

\subsection{Comparison with State-of-the-Art DNN Approaches}\label{sec:comp}
Table~\ref{table3} provides a comparison between our proposed approach $\PH$ and the available methodologies, which shows the superiority of the $\EM$ architecture over the experimental results obtained from the state-of-the-art researches~\cite{WeiNet,YuNet,DingNet,ZhaiNet,1_Ninapro,AtzoriNet}. This comparison was evaluated based on the same settings for the $\DB$ dataset and its sub-exercises, i.e., $\EF$, $\ES$, and $\ET$. Descriptions of this database are provided in sub-section~\ref{sec:dataset}. As stated previously, according to the recommendations in~\cite{24}, the window size should be less than $300$ms to meet the acceptable delay time for myoelectric control systems. Therefore, in this study, we segmented sEMG signals with three windows, i.e, $200$ms, $150$ms, and $100$ms,  to fulfill the mentioned limitation. As shown in Table~\ref{table3}, our proposed approach $\PH$ achieved higher accuracy than the existing methodologies when evaluated based on $\DB$, $\EF$, $\ES$, $\ET$, and different time window sizes. More specifically, we compared the proposed architecture with both advanced DNNs and classical ML approaches.

\begin{table*}[t!]
\centering
\renewcommand\arraystretch{1.8}
\caption{\small Comparison between our methodology ($\PH$) and previous works~\cite{WeiNet,YuNet,DingNet,ZhaiNet,1_Ninapro,AtzoriNet}.\label{table3}}
{\begin{tabular}{c  c  c c c}
\hline
\hline
\multicolumn{1}{c}{\multirow{2}[2]{*}{\textbf{Method}}}
& \multicolumn{1}{c}{\multirow{2}[2]{*}{\textbf{Database}}}
& \multicolumn{3}{c}{\textbf{Window size}}
\\
\cline{3-5}
&
& \multicolumn{1}{c}{\multirow{1}[2]{*}{\textbf{200ms}}}
& \multicolumn{1}{c}{\multirow{1}[2]{*}{\textbf{150ms}}}
& \multicolumn{1}{c}{\multirow{1}[2]{*}{\textbf{100ms}}}
\\
\cline{1-5}
CNN~\cite{WeiNet}
& $\DB$ 
& $83.70$
& $82.70$
& $81.10$
\\
Attention-based Hybrid CNN-RNN~\cite{YuNet}
& $\DB$ 
& $82.20$
& -
& -
\\
CNN~\cite{DingNet}
& $\DB$ 
& $78.86$
& -
& -
\\
CNN~\cite{ZhaiNet}
& $\DB$ 
& $78.71$
& -
& -
\\
CNN~\cite{AtzoriNet}
& $\DB$ 
& -
& $60.27$
& -
\\
SVM~\cite{ZhaiNet}
& $\DB$ 
& $77.44$
& -
& -
\\
RF~\cite{1_Ninapro}
& $\DB$ 
& $75.27$
& -
& -
\\
RF~\cite{DB5}
& $\DB$ 
& $72.25$
& -
& -
\\
$\textbf{\PH}$
& $\DB$ 
&$\textbf{86.18}$
&$\textbf{85.43}$
&$\textbf{84.13}$
\\
\hline
CNN~\cite{ZhaiNet}
& $\EF$ 
& $82.22$
& -
& -
\\
CNN~\cite{DingNet}
& $\EF$ 
& $83.79$
& -
& -
\\
SVM~\cite{ZhaiNet}
& $\EF$
& $81.07$
& -
& -
\\
$\textbf{\PH}$
& $\EF$ 
&$\textbf{88.91}$
&$\textbf{88.14}$
&-
\\
\hline
CNN~\cite{ZhaiNet}
& $\ES$ 
& $72.62$
& -
& -
\\
SVM~\cite{ZhaiNet}
& $\ES$
& $71.08$
& -
& -
\\
$\textbf{\PH}$
& $\ES$ 
&$\textbf{81.44}$
&$\textbf{79.99}$
&-
\\
\hline
CNN~\cite{ZhaiNet}
& $\ET$ 
& $89.54$
& -
& -
\\
SVM~\cite{ZhaiNet}
& $\ET$
& $88.56$
& -
& -
\\
$\textbf{\PH}$
& $\ET$ 
&$\textbf{93.84}$
&$\textbf{93.58}$
&-
\\
\hline
\hline
\end{tabular}}
\end{table*}

For example, Reference~\cite{AtzoriNet} showed the average classification accuracy obtained using all the classical methods such as SVM, RF, K-Nearest Neighbors (K-NN), and LDA on the $\DB$ dataset is $60.28\%$. They achieved the highest gesture recognition accuracy for RF which is $75.27\%$. Moreover, in Reference~\cite{ZhaiNet}, they achieved the recognition accuracy $77.44\%$ using SVM over all the movements. In addition, the recognition accuracy of $72.25\%$ is reported in Reference~\cite{DB5} for the RF classifier. For DNN architectures, on the other hand, the best detection accuracy is reported in Reference~\cite{WeiNet} using CNN, which is $83.70\%$. As shown in Table~\ref{table3}, for window size of $200$ms,  our proposed architecture achieved $86.18\%$ classification accuracy which is $2.48\%$ higher than the state-of-the-art DNN approach and $8.74\%$ higher than state-of-the-art classical ML method. Moreover, it can be observed that for other window sizes, the classification accuracy of our proposed approach achieved better gesture recognition performances than its counterparts. For example, when the  window size is set to $100$ms, our proposed approach $\PH$ was able to achieve gesture recognition accuracy of $84.13\%$, but using the proposed approach of~\cite{WeiNet}, accuracy of $81.1\%$ is achieved. It should be noted that the accuracy of $84.13\%$ obtained by $\PH$ with a window size of $100$ms is still higher than the case where the window size in Reference~\cite{WeiNet} has doubled, i.e., $200$ms. We also evaluated and compared our proposed method for $\EF$, $\ES$, and $\ET$ with the previous studies~\cite{ZhaiNet, DingNet}, which demonstrates the superiority of our hybrid transformer-based framework.

\subsection{Ablation Study}\label{sec:abl}
\begin{figure}[t!]
\centering
\includegraphics[scale=.4]{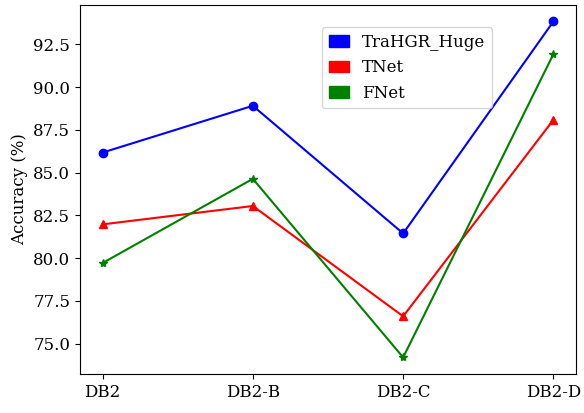}
\vspace{-.1in}
\caption{\small The accuracy for $\PH$, $\vitF$, and $\vitS$ when they are trained simultaneously  for $\DB$ and its sub-exercises, $\EF$, $\ES$, and $\ET$.}\label{fig:out}
\end{figure}
\begin{figure}[t!]
\centering
\includegraphics[scale=.4]{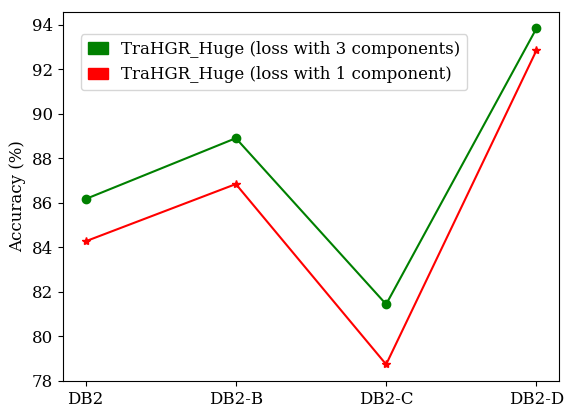}
\vspace{-.1in}
\caption{\small Results of the ablation study on loss functions with $\PH$ model which is trained by Eq.~\ref{eq:loss} (green) and Eq.~\ref{eq:loss1} (red) evaluated on $\DB$, $\EF$, $\ES$, and $\ET$.}\label{fig:loss}
\end{figure}
For the proposed hybrid architectures such as $\PH$, $\PL$, and $\PB$, the classification accuracy is calculated using the prediction values $\y$ obtained from Eq.~\eqref{eq:out}. To show that our proposed architecture based on a developed hybrid strategy has great potentials for improving gesture recognition accuracy, we also calculated the other two accuracies, i.e., $\y_{\vitF}$ or $\y_{\vitS}$, based on the Eq.~\eqref{eq:out1}. More specifically, we trained the hybrid architectures by computing the loss function in Eq.~\eqref{eq:loss}. However,  output $\y$ is used to calculate the accuracy, reported in this paper. Here, in Fig.~\ref{fig:out}, it is shown that the accuracy obtained using the $\y$ is better than those calculated using the $\y_{\vitF}$ or $\y_{\vitS}$ for $\DB$ and its sub-exercises. In particular, from Fig.~\ref{fig:out}, it can be observed that the hybrid architecture takes advantage of two parallel paths and improved the recognition accuracy. 

\subsubsection{Evaluation of Different Loss Functions}\label{sec:loss}
As described in sub-section~\ref{sec:loss1}, our proposed hybrid architectures' parameters are learned by optimizing the loss function $\mathcal{L}$, which consists of three components. To demonstrate the advantage of training our proposed hybrid architecture  with loss function $\mathcal{L}$ defined in Eq.~\eqref{eq:loss}, we evaluated performance of $\PH$ when the loss function $\mathcal{L}$ has only one component as follows
\begin{eqnarray}
\mathcal{L} = \mathcal{L}_{\EM} \label{eq:loss1}.
\end{eqnarray}
Fig.~\ref{fig:loss} shows the performance of $\PH$ in $\DB$ and its sub-exercises for two different loss functions. We can see that training $\PH$ with a loss function with three components (Eq.~\eqref{eq:loss}) improves the results compared to the case where loss function has only one component (Eq.~\eqref{eq:loss1}). 

\vspace{-.1in}
\section{Conclusion}  \label{sec:con}
In this paper, we proposed a hybrid architecture based on transformers for the task of hand gesture recognition. We have shown that the proposed hybrid  architecture, referred to as the $\EM$ framework, could augment the power of model discrimination in different scenarios for various exercises. Moreover, we investigated the ability of transformers for sEMG-based hand gesture recognition as they have revolutionized other fields such as NLP, CV, and speech recognition. In this regard, we compared $\EM$ results with traditional ML approaches and DNN-based techniques and demonstrated the outstanding performance of the proposed architecture.  A potential direction for future research is to use the proposed transformer-based architecture to develop an adaptive learning method with focus on increasing the robustness of sEMG classifiers and improving inter-subject accuracy will be an interesting direction for future research. Another direction for future research is to use and extend the proposed transformer-based hybrid architecture to other machine learning fields.

\bibliographystyle{IEEEbib}

\end{document}